\documentclass[fleqn,twoside]{article}
\usepackage{espcrc2}

\usepackage{graphicx}
\usepackage[figuresright]{rotating}

\newcommand{\openone}{\lea{\it 
vev}mode\hbox{\small1\kern-4.2pt\normalsize1}}
\newcommand{\rr}[4]{#1, {\it #2 \/}{\bf #3} #4}

\newcommand{\al}{\alpha}
\newcommand{\alef}{\al_{eff}'}

\newcommand{\sg}{\sigma}

\newcommand{\alp}{\alpha'}

\newcommand{\eq}{\[}
\newcommand{\f}[2]{\frac{#1}{#2}}
\newcommand{\eqx}{\]}
\newcommand{\eqn}{\begin{eqnarray}}
\newcommand{\eqnx}{\end{eqnarray}}

\newcommand{\DD}{{\cal D}}

\newcommand{\cor}[1]{\left\langle{#1}\right\rangle}
\renewcommand{\AA}{{\cal A}}
\renewcommand{\DD}{{\cal D}}

\newcommand{\AmS}{{\protect\the\textfont2
  A\kern-.1667em\lower.5ex\hbox{M}\kern-.125emS}}

\title{Regge amplitudes from AdS/CFT duality}

\author{R. Peschanski\ \address{CEA/DSM/SPhT,Unit\'e de recherche 
associ\'ee 
au CNRS, \\
CE-Saclay, F-91191 Gif-sur-Yvette Cedex,France; \\email: 
pesch@spht.saclay.cea.fr}}
       
\begin{document}

\begin{abstract}

\vspace{1pc}
String theory has long ago  been initiated by the quest for a 
theoretical 
explanation of     the observed high-energy ``Regge behaviour'' of  
strong 
interaction amplitudes, but  this  35-years-old puzzle is still unsoved.  
We 
discuss how modern tools like the  AdS/CFT correspondence give a new 
insight 
on the problem.
\end{abstract}

\maketitle

\section{Introduction}

\begin{figure}
\includegraphics[width=17.5pc]{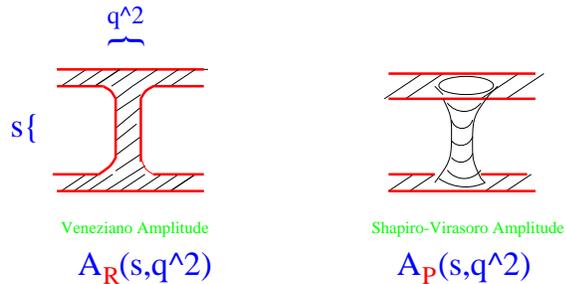}
\vspace{-18pt}
\caption{Reggeon and Pomeron String Amplitudes.}
\label{1}
\end{figure}

It is well-known that string theory started from the proposal of  
scattering 
amplitudes which may grasp the two major 
structures of soft interaction phenomenology in a condensed form : 
resonances 
and Regge poles. Two  types of amplitudes were proposed 
for 
four-point amplitudes, see Fig.\ref{1}. The Veneziano amplitude 
corresponds to {\it Reggeon}  
exchanges with non-vacuum quantum 
numbers  and the Shapiro-Virasoro amplitude  corresponds to {\it 
Pomeron} 
exchange with vacuum quantum numbers. As soon after demonstrated, they 
correspond to respectively open and closed bosonic string 
theory 
amplitudes at tree-level. 

However, despite many efforts during years, no widely recognized  
progress 
have 
been done in the string theory of strong interaction amplitudes, and 
after 
the 
discovery of QCD as the gauge field theory of quarks and gluons, there 
remained 
little place for it. Indeed, major theoretical obstacles have been 
raised, 
for 
instance:
\begin{itemize}
\vspace{-.2cm}\item The conformal anomaly of string theories in 
Minkowski 
$D$-dimensional 
space leads to the limitation  $D=26,10$ for  bosonic and 
super strings. 
\vspace{-.3cm}\item Zero mass gauge and  gravitational fields appear in 
the  
string spectra 
of 
 asymptotic states, if not tachyonic states.
\end{itemize}
\vspace{-.2cm}
To these (non exhaustive list of) difficulties, new questions  have to 
be 
added  
after  the discovery of QCD, for instance: 
\vspace{-.3cm}
\begin{itemize}
\vspace{-.3cm}\item Where  are ``hard'' interactions recovered in a 
 string 
theory framework?
\vspace{-.3cm}\item Can we elaborate a suitable string Theory which 
could  
coherently describe 
the  properties  of gauge fields?
\end{itemize}
To these pending questions, the recently proposed AdS/CFT correspondence 
between 
certain string and gauge field theories may 
give new and reliable answers. Among these questions, the understanding 
of  
Regge amplitudes in terms of the AdS/CFT duality have been the subject 
of 
an 
approach which I will now describe.

\section{String/Gauge fields Duality}

\begin{figure}
\includegraphics[width=17.5pc]{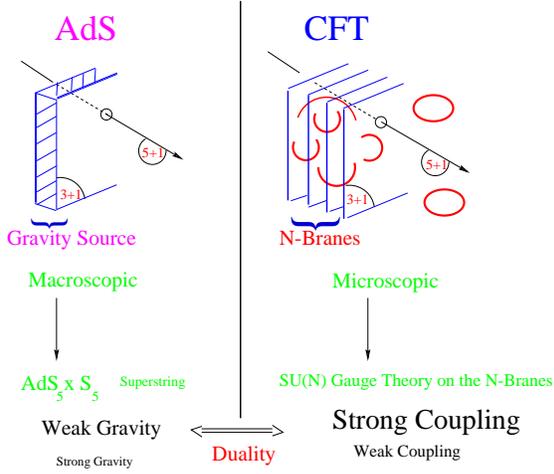}
\vspace{-18pt}
\caption{AdS$_5$/CFT$_4$ duality correspondence.}
\label{2}
\end{figure}

The AdS/CFT correspondence \cite{11}  has many interesting formal and 
physical facets. Concerning 
the 
aspects which are of interest for our problem, it allows one to find 
relations 
between gauge field theories at strong coupling and string gravity at 
weak 
coupling  in the limit of large number of colours ($N_c\!\to\! \infty$), 
see 
Fig.\ref{2}. It can be examined  quite precisely in  the 
AdS$_5$/CFT$_4$ 
case which conformal field theory  corresponds to $SU(N)$ gauge 
theory 
with ${\cal N} \!=\!4$ supersymmetries. 

Some existing extensions to other gauge theories with  broken conformal 
symmetry
with less or no supersymmetries will be valuable for our approach, since 
they 
lead to confining gauge theories  which 
are 
more similar to  QCD. Indeed, one important question is  to examine to 
what 
extent confinement plays a r\^ole in the Reggeization of amplitudes.

However,  note that the 
appropriate 
string gravity dual of QCD has not yet been identified, and thus we are 
forced 
to restrict for the moment our  use of AdS/CFT correspondence to 
features 
which 
are 
expected to be a general feature of  confining theories duals.

Let us schematically recall the canonical derivation of the AdS$_5$ 
background. One starts from  the (super)gravity classical 
solution of a system of $N\ D_3$-branes in a $10-D$ space of the (type 
IIB) 
superstrings. 
The metrics solution of the (super)Einstein equations read
\eq
\label{super}
ds^2=f^{\!-\!1/2} (\!-\!dt^2\!+\!\sum_{1\!-\!3}dx_i^2) 
\!+\!f^{1/2}(dr^2\!+\!r^2d\Omega_5) 
\ ,
\eqx
where the first four coordinates are on the brane and $r$ corresponds to 
the  
coordinate along the normal.
\eq
f=1+\frac {R^4}{r^4}\ ;\ \ \ \ \ R=4\pi g^2_{YM}\alpha'^2 {N} \ ,
\eqx
and $g^2_{YM}{N}$ is the `t Hooft-Yang-Mills coupling and $\alpha'$ the 
string 
tension. Then one introduce the  ``Maldacena limit'', where 
one sits near-by the branes  while in the same time going to weak string 
coupling limit. the space-time is thus distorted due to the (super) 
gravitational field of the branes. One writes 
\eq
\frac {\alpha'(\to 0)} {r (\to 0)}\to z\ ,\ R\ fixed\ \Rightarrow 
g^2_{YM} 
{N}\sim \f{1}{\alpha'^2}\to \infty \ .
\eqx
By reorganizing the two parts of the metrics one obtains
\eq
ds^2={ \frac 1{z^2} (-dt^2+\sum_{1-3}dx_i^2+ dz^2)} +  {R^2 
d\Omega_5}\ ,
\eqx
which corresponds to the 
{AdS$_5$} $\times \  {S_5}$ background structure,  ${S_5}$ being  
the 5-sphere. More detailed analysis shows that the isometry group of 
the  
5-sphere is the geometrical dual of the ${\cal N}\! =\!4$ 
supersymmetries.

In order to illustrate the way  how one formulates the AdS/CFT 
correspondence, 
let us consider the example of  the vacuum expectation value ({\it vev}) 
of 
Wilson lines in a  configuration parallel
 to the time direction of the branes. This configuration
allows 
a   determination of the 
potential between colour charges \cite{12}.

\begin{figure}
\includegraphics[width=17.5pc]{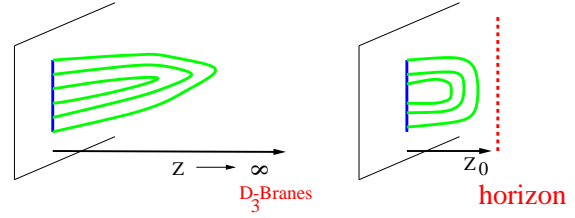}
\vspace{-18pt}
\caption{Exemple of minimal surfaces with Wilson lines boundary.}
\label{3}
\end{figure}
One writes
\[
\langle e^{iP\int_C\vec A\cdot\vec dl}\rangle 
\!=\!\!\int_{\Sigma}\!e^{-\frac 
{Area({\Sigma})}{\alpha'}}\!\!\approx e^{-\frac  
{Area_{min}}{\alpha'}}\times 
Fluct.\ ,
\]
where $C$ is the Wilson line contour near the $D_3$ branes  and $\Sigma$ 
the 
surface in $AdS$-space with $C$ as the boundary, see Fig.\ref{3}. The 
minimal 
area 
approximation is  the {\it vev} evaluation classical $\alpha'\to 0$ 
limit which 
can 
eventually be improved  by calculating the  fluctuation 
determinant  
around the minimal surface. In  Fig.\ref{3}, we have sketched Two  
cases: the $AdS_5$
``conformal'' 
one and a 
a 
confining 
case, $AdS_{BH},$ where a black-hole (BH) in the AdS bulk determines a 
characteristic horizon scale $R_0$  breaking  conformal invariance (see 
Witten \cite{11}).

The {\it vev} results can be summed up as follows:
\eqn
AdS_5: \langle Wilson\ Lines\rangle\!\!\!\!&=&\!\!\!\!e^{TV(L)}\sim 
e^{\#T/L}\nonumber \\ AdS_{BH}: \langle Wilson\ 
Lines\rangle\!\!\!\!&=&\!\!\!\!e^{TV(L)}\sim e^{\#TL/R_0^2}\ , \nonumber 
\eqnx 
where, the potential behaviour is as expected for respectively conformal 
(perimeter law) and confining (area law) cases.   Note that  there is   
interesting  information  in the coupling dependent numbers 
here denoted by  $\#$ .

\section{Supergravity Duals of Scattering Amplitudes}

Interestingly enough, high energy amplitudes in gauge field theories can 
be 
related to other related configurations of 
minimal surfaces \cite{13}. At high energy, fast moving colour sources 
propagate 
along
linear trajectories in coordinate space thanks to the eikonale 
approximation. An analytic 
continuation 
from Minkowski to Euclidean ${\cal R}^4$ space allows one to find a  
geometrical interpretation in terms of a well-defined minimal 
surface 
problem. Let us 
consider for illustration different applications. 

\subsection{``Quark'' elastic scattering}

\begin{figure}
\includegraphics[width=8pc]{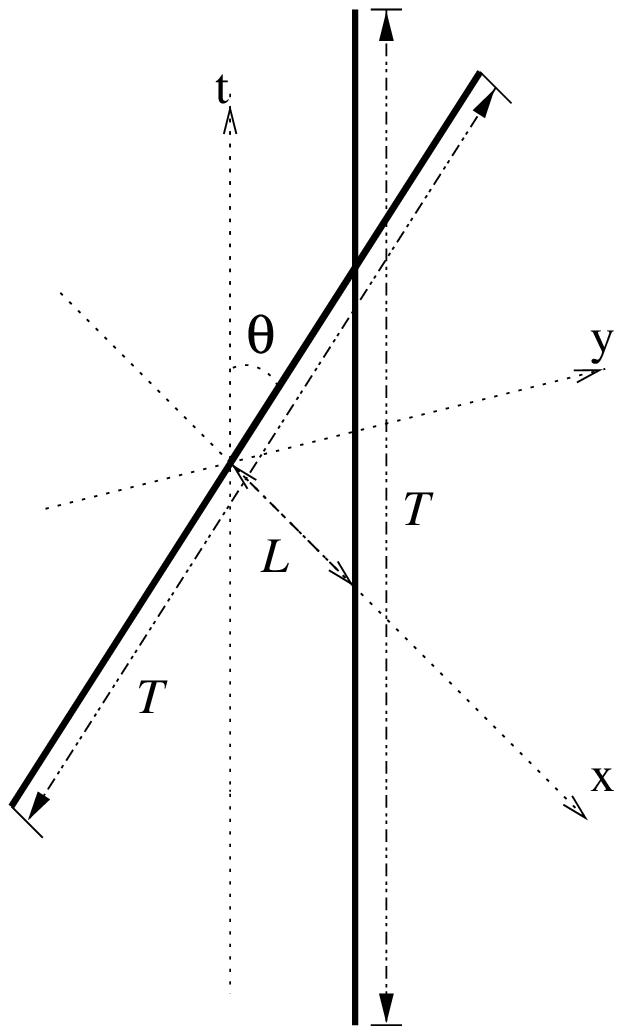}
\includegraphics[width=8.5pc]{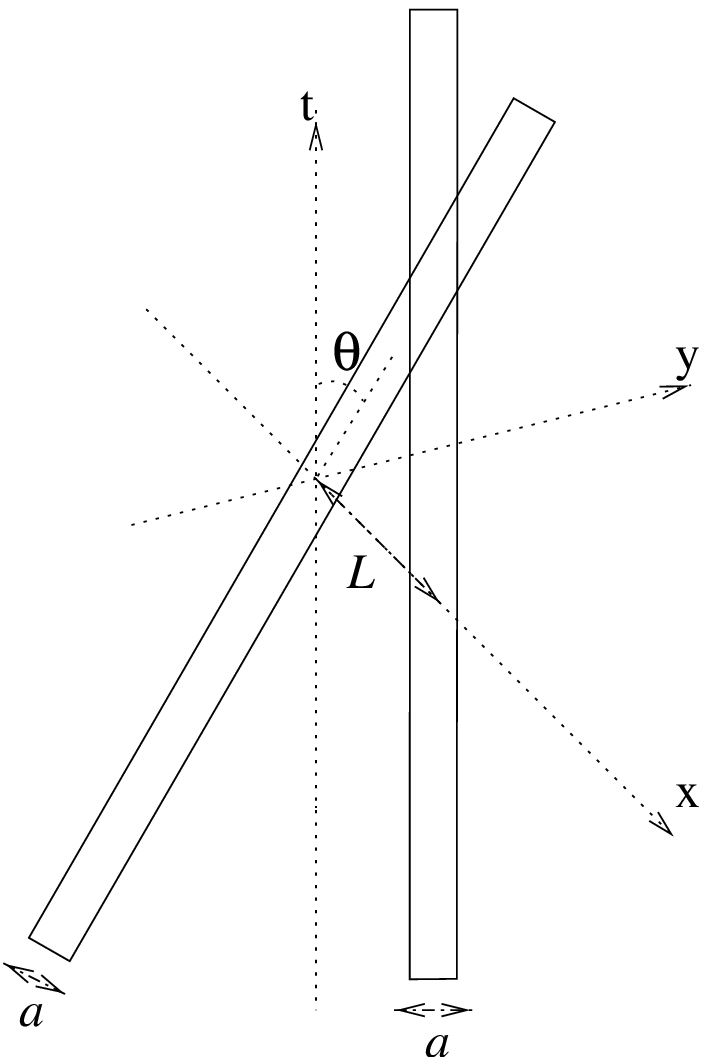}
\vspace{-18pt}
\caption{Wilson lines for  ``quark'' and dipole elastic scattering in 
${\cal 
R}^4.$}
\label{4}
\end{figure}

Calling ``(anti)quarks'' the colour sources in the (anti)fundamental 
representation of SU(N), the high-energy elastic quark-(anti)quark
amplitude can be written \cite{14}
\[
A(s,q^2)=2is\int d\vec l\ e^{i\vec q\cdot \vec l}\ \langle W_1 
W_2\rangle 
_{L=|\vec l|}^{\chi=\log s/m^2}\ ,
\]
where $\vec l$ is the impact parameter between the two trajectories, 
conjugated to the momentum transfer $\vec q,$  $\chi$ the total 
rapidity 
interval. Performing an analytic 
continuation to the Euclidean space:
\[
\chi \to i\theta\ \ \ ;\ \ t_{Mink}\to -it_{Eucl}\ ,
\]
the Wilson line {\it vev} can be expressed as a minimal surface problem 
whose 
boundaries are two straight lines with an angle $\theta$ in ${\cal 
R}^4,$ 
see 
Fig.\ref{4}.  In flat space, with the same boundary conditions,  the 
minimal surface is 
the 
helicoid. One thus  realizes that the problem can be formulated as a 
minimal surface problem whose  mathematical formal solution   is a 
{\it generalized  helicoidal} manifold embedded in Euclidean AdS Spaces.
\subsection{Dipole elastic scattering}

For elastic  scattering of colourless states, it is interesting to 
consider 
QCD dipoles,  which  are known to be  a good toy model. Their  
propagation 
in 
coordinate space within the eikonale approximation can be 
represented by elongated Wilson loops near both right and left moving 
light-cone 
directions. Using the same analytic continuation framework, one has to 
compute the Wilson loop correlator in the configuration displayed in 
Fig.\ref{4}.

\begin{figure}
\includegraphics[width=17.5pc]{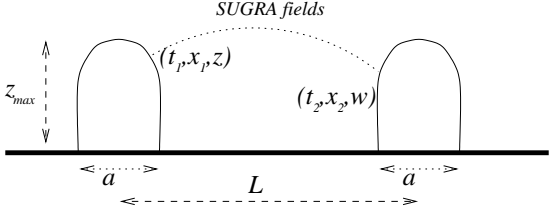}
\vspace{-18pt}
\caption{Dipole scattering at large impact parameter.}
\label{5}
\end{figure}

The minimal area solution with the corresponding  boundary conditions  
is difficult to find in analytic form, necessary for the continuation to 
Minkowski space. Some approximation schemes may be used \cite{12}. A 
quite 
general and intringuing feature is the existence of a  geometrical  
transition 
between small and  large impact parameter, corresponding to the 
realization 
of 
disconnected minimal surfaces, see Fig.\ref{5}.

\subsection{Dipole  inelastic scattering}

\begin{figure}
\includegraphics[width=15pc]{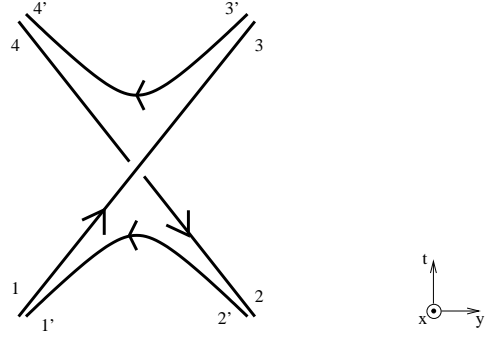}
\vspace{-18pt}
\caption{Wilson lines for inelastic dipole scattering}
\label{6}
\end{figure}

The application of AdS/CFT correspondence for the two previous exemples 
is 
not 
so easy, even if partial results are encouraging. For ``quark'' elastic 
scattering, an infra-red time-like cut-off is to be introduced due to 
the 
colour 
charges of the quarks which implies a regularization scheme and a 
complication 
of the geometrical aspects.  For dipole  
elastic scattering, there is no need for a cut-off but the geometry of 
the 
minimal surface is complicated. Inelastic scattering of dipoles allows 
one 
to 
circumvent these difficulties. Indeed, the helicoidal geometry remains  
valid 
due 
to the eikonale approximation for the ``spectator quarks''
while the ``exchanged quarks'' define a trajectory drawn on the 
helicoid, see 
Fig.\ref{6}. This trajectory plays the r\^ole of a dynamical time-like  
cut-off which takes part in the 
minimization 
procedure.

Adopting the ``world-line'' path integral scheme of Feynman \cite{15}, 
one may write 
the 
inelastic amplitude in terms of a Wilson loop {\it vev}:
\[
\int \DD\tau\, \cor{W(1\!\to\! 3'\! \to \!4'\! \to \!2' \!\to \!1')}_
{\AA,[\tau]} 
\ e^{ -2m {\cal L}[\tau]} \ ,
\]
where $\tau$ parametrizes the boundary trajectories and ${\cal L}$ is 
their 
total length. Using the AdS-CFT correspondence in the same framework as 
previously,  
one may formally integrate   over  the gauge degrees of freedom and 
write 
\[
\cor{W(1\!\!\to\!\! 3'\!\! \to \!\!4'\! \!\to\! \!2' \!\!\to 
\!\!1')}_{\AA,[\tau]}\!\! =\! 
e^{-\!\f{Area[\tau]}{2\pi\alpha'} } \times Fluct.
\]
Note that the remaining minimization in $\tau$ runs now on both  the 
area  and its 
boundary.

\section{Reggeization from the geometry of its AdS dual}

To give a practical exemple of calculation  we focus on the 
configuration 
of Wilson lines of Fig.\ref{6} in the context  of a confining theory, 
dual to the $AdS_{BH}$ case with a 
limiting 
horizon, see Fig.\ref{3}.  The  AdS$_{BH}$ metrics, as well as the 
canonical 
AdS$_5$ one, are characterized by a singularity at $z=0$ which 
implies a 
rapid growth in the  $z$ direction 
towards the D$_3$ branes, then stopped near the horizon at $z_0.$ Thus, 
to a good approximation, and for large enough
impact 
parameter (compared to the horizon distance), the main contribution to 
the 
minimal area is from the  metrics in the bulk near $z_0$ which is nearly
flat. Hence, near $z_0,$ the relevant minimal area can be  drawn on a
classical helicoid .
Parametrizing  the helicoid:
\eqn
t &=& \tau \cos {\theta\sg}/{L}\nonumber\\ 
y &=& \tau \sin {\theta\sg}/{L}\nonumber\\
x &=& \sg\nonumber\\  
z &\sim& z_0\nonumber
\eqnx
the solution of the amplitude boils down to an Euler-Lagrange  
minimization 
over $\tau,$ namely
\eqn
A_{\cal R}(s,L^2)&\propto&\f{1}{s}\lim_{\alp\to 0}\int \DD\tau\,  
e^{-\f{1}{2\pi\alpha'} Area[\tau]}
\times \nonumber\\
&\times& e^{-2m {\cal L}[\tau]}\times Fluct. \nonumber\ ,
\eqnx
where $Area[\tau]$ is the section of an helicoid bounded by the quark 
trajectories  having total length ${\cal L}[\tau].$ 
Note that the kinematical factor ${1}/{s}$  in front comes from an  
infinite-dimensional fermionic spin factor along the quark trajectories 
corresponding to a bosonized
representation \cite{15}. Its non-trivial calculation makes  
use 
of  
the 
3-dimensional embedding of the quark trajectories on the helicoid 
surface \cite{12}.

After some technical steps and analytic continuation back to Minkowski 
space, 
the 
resulting amplitude reads:
\eq
A_{\cal R}(s,q^2)=\int d\vec l\ e^{i\vec q\cdot \vec l}\ e^{-\f{ 
L^2}{4\alef 
\chi}}\propto  s^{-\alef q^2}\ ,
\eqx
corresponding to a linear Regge trajectory with intercept $0$ and 
slope $\alef$ related to the quark potential calculated within the same 
AdS/CFT 
framework. 

\begin{figure}
\includegraphics[width=17.5pc]{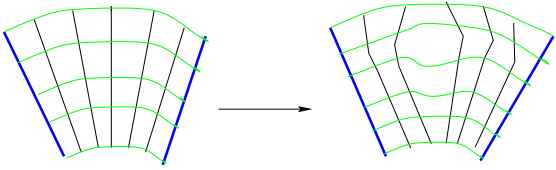}
\vspace{-18pt}
\caption{Fluctuations around the minimal surface}
\label{7}
\end{figure}

A semi-classical correction   comes from the fluctuations near the 
minimal 
surface sketched 
in Fig.\ref{7}. It  can be shown to be intimately related to a 
contribution to the  quark potential similar to the well-known L\"uscher 
term \cite{16}. 
One finally finds
\eq
 A_{\cal R}(s,q^2)\propto s^{\f{n_\perp}{24}-\alef 
q^2}\ ,
\eqx
where $n_\perp$ is the number of transverse zero-modes (zero-mass 
transverse 
excitations of the string) 
in the AdS dual theory. It gives a shift in the intercept of the reggeon 
trajectory. 

The value of this intercept 
depends on the particular realization of 
the AdS/CFT duality. In known examples (see  \cite {12}) it is quoted to 
be $n_\perp\!=\!7\ or\ 8,$ , to be compared with the 
ordinary   L\"uscher term having  $n_\perp=2.$. 
Note 
that fermionic d.o.f. are not expected to remain massless in an AdS 
background, and thus to give contributions with opposite sign. 

It is worth comparing the  resulting {\it inelastic} dipole amplitude 
with the one  obtained for dipole {\it elastic} 
scattering in the same framework. 
With  approximations required by to the non-trivial geometry of 
dipole 
loops  one finds:
\eq
A_{\cal P}(s,q^2) \sim s^{1+\f{n_\perp}{96}-\f\alef 4 
q^2}\ ,
\eqx
with the same values of $\alef $ and $n_\perp.$  Reggeon 
and 
Pomeron Regge trajectories are thus linear and related to
the quark potential. Note the factor $1/4$ in the exponent of $A_{\cal 
P}$ with respect to $A_{\cal R}$ which has some   consistency with 
actual amplitudes, since it gives   phenomenologically consistent 
Pomeron and (average) dominant Reggeon trajectories \cite{12}. 

\section{Conclusion: Reggeization and confinement}

Lattice calculations, which is the only presently known way to evaluate 
directly QCD observables at strong coupling, are not able to compute 
high-energy amplitudes. hence, an interesting output of the application 
of 
AdS/CFT correspondence to high energy amplitudes at strong coupling is 
to 
discuss the relation between Reggeization and confinement, using the 
description in the dual theory.

When comparing  AdS$_5$  duality - which corresponds to a conformal, 
non-confining gauge theory - with   AdS$_{BH}$  duality, which leads 
to 
reggeization, the difference ultimately comes from the different metrics 
in 
the bulk and hence from the minimal surfaces for the same boudary 
conditions. Taking into account  their different geometry, see e.g.  
Fig.\ref{3}, one expects after analytic continuation and in the 
large 
energy ($\chi\to \infty$) limit:
\eq
Area^{AdS}_{min}\sim 
\lim_{\chi\to \infty} \f{L}{L/\chi}\ ; \ Area^{BH}_{min}\sim 
\lim_{\chi\to 
\infty} 
L\times\f{L}{\chi}\ .
\eqx 
The AdS$_5$  case leads to a $L$-invariant value  and, 
after 
Fourier transformation, to a high-energy amplitude with a $q^2$ 
independent 
energy exponent (or flat Regge trajectory). On the other hand, the 
AdS$_{BH}$ 
case  leads to a linear Regge trajectory 
after 
Fourier transformation. 
For the AdS$_{BH}$ case this rough expectation can be verified by
an explicit calculation. Hence confinement appears as an essential 
ingredient 
for the reggeized structure of high-energy amplitudes. We expect this 
result 
 not to be dependent on the precise geometrical 
AdS$_{BH}$ setting and thus to indicate a quite general property of 
confining 
theories.

As a conclusion, let us discuss  
the   
list of problems  
for which the AdS/CFT framework give new insights on the  35-years-old 
puzzle of high-energy amplitudes at strong gauge coupling\footnote{A different 
approach has been proposed in Ref.\cite{17}.}. As usual this 
also adds some new 
problems 
in this context!
Among the new insights:
\begin{itemize}
\vspace{-.1cm}\item {\it String dimensionality:} $D=4+1+5:$ 
Extra-dimensions 
play an important dynamical r\^ole. 
\vspace{-.3cm}\item {\it Gravitation:} It is  decoupled. 
The (classical) Pomeron intercept is  $1+\epsilon$ instead of $2$ for 
the graviton.  $\epsilon$ is related to a L\" usher term.
\vspace{-.5cm}\item {\it Regge trajectories:} They come out  linear, 
with 
slopes and intercepts related to the quark potential including a  
L\" uscher 
term.
\vspace{-.5cm}\end{itemize}\vspace{-.1cm}
Among the problems: 
\begin{itemize}
\vspace{-.1cm}\item {\it High-energy phenomenology:} Many aspects, like 
the 
Flavor/Spin dependences, remain  to be studied.  
\vspace{-.3cm}\item {\it Approximations:} The dual gauge theory is not 
specified, and the exact minimal surface in the bulk metrics to be 
determined. 
\vspace{-.3cm}\item {\it Dual of QCD?} In the present framework, the 
confining  scale $R_0$ has no relation with  $\Lambda_{QCD}.$
\vspace{-.3cm}\item {\it Unitarity:}  A more complete investigation  
requires the study of multi-leg amplitudes.
\vspace{-.6cm}\item {\it Deeper general problems:} The formulation of 
string 
theory in AdS backgrounds and last but not least, a proof of the AdS/CFT
conjecture.
\vspace{-.4cm}
\end{itemize}

\section{Acknowledgement} 
The whole approach of  high-energy amplitudes described in this review 
paper 
comes from a collaboration with Romuald Janik from Cracow University.

\end{document}